\documentstyle{article}
\font\tenbf=cmbx10
\font\tenrm=cmr10

 1
 1
 1

\font\ninerm=cmr9
\font\nineit=cmti9

\font\eightrm=cmr8

\renewenvironment{thebibliography}[1]
 { \tenrm
   \begin{list}{\arabic{enumi}.}
    {\usecounter{enumi} \setlength{\parsep}{0pt}
     \setlength{\itemsep}{2pt} \settowidth{\labelwidth}{#1.}
     \sloppy
    }}{\end{list}}

\begin{document}
\begin{center}{{\tenbf
              STRANGE DIBARYONS IN THE SKYRME MODEL
\footnote{\eightrm\baselineskip=11pt
The work supported in part by Russian Fund for Fundamental Researches,
grant 95-02-03868a}
\\}
\vglue 0.7cm
{\ninerm V. B. KOPELIOVICH \\}
{\nineit Institute for Nuclear Research of the Russian Academy of
Sciences,\\ 60th October Anniversary Prospect 7A, Moscow 117312, Russia\\}
\vglue 0.5cm}
{\begin{minipage}{4.2truein}
\eightrm
\baselineskip=10pt
\noindent
The phenomenological consequences of the existence of different local
minima in the SU(3) configuration space of B=2 skyrmions are 
discussed.
\end{minipage}}
\end{center}
\vglue 0.7cm
\baselineskip=14pt
\tenrm

1. Since the prediction of the existence of $H$-dibaryon in the framework
of MIT quark-bag model $^1$ and confirmation of this prediction also 
in the framework of Skyrme model $^2$ many efforts have been done
to study theoretical predictions for the spectrum of baryonic systems
within different approaches. The chiral soliton approach is of special
interest because it provides unconventional point of view at the
baryonic systems and/or nuclear fragments. Bound states of chiral 
solitons or skyrmions appear as objects where the baryons individuality
is lost and can be reconstructed when quantum effects are taken into
account $^{2-5}$.

Till now three different types of dibaryons have been established within
the chiral soliton approach.
Historically the first one was obtained as $SO(3)$ soliton $^2$.
The state with the lowest possible winding number is $B=2$ hedgehog
being interpreted as $H$-dibaryon,
the state with the azimuthal winding $n=2$ has $B=4$ and the torus-like 
form of mass and $B$-
number distributions $^6$. It is bound relative to the 
decay into two $B=2$ hedgehogs $^6$. Recently the bound state
of two $H$-dibaryons was obtained also within the framework of quark-
cluster model (T.Sakai, the talk at this Workshop). 
As it was shown in $^7$ the $H$-particle can be unbound when 
Casimir energies (CE) of solitons are taken into account.

It should be noted that within the framework of the chiral soliton
models the $H$-particle is a rather small object, $\sqrt{R_H^2} \sim
0.5-0.6 Fm$ $^7$, see also $^4$. So, it can be considerably smaller
than the deuteron, and this may be the reason why the $H$-particle was
not observed experimentally up till now: in theoretical estimates of
the cross sections it was assumed often that the $H$-particle is
an extended object, similar to the deuteron. \\

2. The second type of dibaryons is obtained by means 
of quantization of bound $SU(2)$-solitons in $SU(3)$ collective 
coordinates space $^{3-5,8}$. The bound state
of skyrmions with $B=2$ possesses generalized axial symmetry and
torus-like distributions of the mass and $B-$number densities. Now
it is checked in several variants of chiral soliton models and
also in the chiral quark-meson model. Therefore, the existence
of $B=2$ torus-like bound skyrmion seems to be firmly established. 

After the zero-modes quantization procedure the $SU(3)$ multiplets 
of dibaryons
appear  with the ratio of strangeness to baryon number 
$S/B$ down to $-3$. The possible $SU(3)$ multiplets which could consist
of minimal number of valence quarks are antidecuplet, $27-$, $35-$ and 
$28-$ plets. The contribution to the energy from rotations
into "strange" direction is the same for all minimal irreps satisfying
the relation $\frac{p+2q}{3}=B$, due to cancellation of second order
Casimir operators of $SU(3)$ and $SU(2)$ groups $^8$. All these
states are bound when contributions linear in $N_c$, the classical
mass, and of the order $N_c^{-1}$, the zero-modes quantum corrections,
are taken into account. However,
after renormalization of masses which is necessary to take into 
account also the $CE$ of the torus (of the order $N_C^0$) and to 
produce the nucleon-nucleon $^1S_0$-scattering state on the right 
place all states with strangeness
different from zero are above thresholds for the strong decays $^8$.
Therefore, it will be very difficult, if possible, to observe such
states experimentally. Another, quite realistic, possibility is that
these states represent virtual bound states in $\Lambda N$, $\Lambda
\Sigma$, etc. systems, similar to the $^1S_0$ $NN$ -scattering state.
In $\Lambda N$ system the virtual state has been seen many years ago
in reactions $pp \rightarrow \Lambda p K^+$, $^9$ , see also the
talk by Y.Fujiwara at the present Workshop. \\ 
 
3. The third type of states
is obtained by means of quantization of strange skyrmion molecules
found recently $^{10}$.
To obtain the strange skyrmion molecule we used the ansatz of the
type

$$        U = U(u,s)  U(u,d)  U(d,s)            \eqno            (1)
$$
where  $U(u,s)$ and $U(d,s)$ describe solitons located in $(u,s)$ 
and $(d,s)$ $SU(2)$ subgroups of $SU(3)$, one of $SU(2)$-matrices, 
 e.g. $U(u,d)$   depends  on two parameters:
$$    U(u,d)= \exp(ia \lambda_2 ) \exp(ib \lambda_3 ) \eqno (2) $$ 
and thus describes the relative local orientation
of these solitons in usual isospace. The configuration considered
depends totally on 8 independent functions of 3 variables. 

To get the $B=2$ molecule we started from two $B=1$ skyrmions in
the optimal attractive orientation at relative distance between 
topological centers close to the optimal one, a bit smaller.
Special algorithm
for minimization of the energy functionals depending on 8 functions
was developed and used $^{10}$. After minimization we obtained the
configuration of molecular type with the binding energy about half
of that of the torus, i.e. about $\sim 70$ $Mev$ for parameters of the
model $F_{\pi}=186$ $Mev$ and $e=4.12$. The attraction between unit 
skyrmions which led to the formation of torus-like configuration
when they were located in the same $SU(2)$ subgroup of $SU(3)$
is not sufficient for this when solitons are located in different
subgroups of $SU(3)$. It is connected with the fact that solitons
located in different $SU(2)$ subgroups interact through only one common
degree of freedom, instead of $3$ degrees, as in the first case.\\

4. The quantization of zero modes of strange skyrmion molecules cannot
be done using the standard procedure, its substantial modification
is necessary $^{11}$. As a result, the quantization condition
established first in $^{12}$ is changed, and for strange
skyrmion molecule we obtained $^{11}$

$$ Y^{min}_R = -1          \eqno(3)  $$

 instead of $Y_R=B$, (we put here the number of colors 
of underlying QCD $N_c=3$), $^{12}$.

The lowest multiplets obtained by means of quantization of
strange skyrmion molecule are octet, decuplet and antidecuplet.
Within the octet the states with strangeness $S=-1$, $-2$ and $-3$
are predicted. They are coupled correspondingly to $\Lambda N$-
$\Sigma N$, $\Lambda\Lambda$-$\Xi N$
or $\Lambda\Sigma$ and $\Lambda\Xi$-$\Sigma\Xi$ channels. 

The mass splittings within multiplets considered are defined, as
usually, by chiral and flavor symmetry breaking mass terms in the 
effective lagrangian. Their contribution to the masses of the states 
in the case of strange skyrmion molecules equals to

$$\delta M = - \frac{1}{4} (F^2_Km^2_K-F^2_{\pi}m^2_{\pi})
 (v_1+v_2-2v_3) <\frac{1}{2} sin^2\nu>    \eqno(4) $$
 
$v_1$, $v_2$ and $v_3$ are real parts of diagonal matrix elements of
unitary matrix $U$, the function $\nu$ parametrizes as usually the 
$\lambda_4$ rotation in the collective coordinates quantization 
procedure and the
average over the wave function of the state should be taken for
$sin^2\nu$. For two interacting undeformed hedgehogs at large
relative distances $v_1+v_2-2v_3 \rightarrow 2(1-cosF)$ where $F$
is the profile function of the hedgehog. Note, that the sign in
$(4)$ is opposite to the sign of analogous term when $(u,d)$
$SU(2)$ soliton is quantized with $SU(3)$ collective coordinates.

The result of calculation depends to some degree on the way of
calculation. We can start with the soliton calculated for all
meson masses equal to the pion mass (flavor symmetric, $FS$-case),
and in this case the central values of masses are $\sim 4.3$,
$4.6$ and $4.8$ $Gev$ for the octet, decuplet and antidecuplet,
the mass splittings are within $130-170$ $Mev$.
This should be compared with the central values of masses of the
octet and decuplet of baryons within the same approach, $2.64$ and
$3.05$ $Gev$ $^{13}$. 
Another possibility is to start with the soliton with the kaon mass
being included into lagrangian
(flavor symmetry broken, $FSB$-case). The static energy of solitons 
are greater in the $FSB$ case, the moments of inertia are smaller, and 
the mass splittings within the $SU(3)$ multiplets are squeezed by a factor
about $\sim 2.7$ in comparison with the $FS$-case.
The central values of masses in $FSB$ case are about $4.33$, $4.9$ 
and $5.4$ $Gev$.
The results of both ways of calculation are close to each other
for the octet of dibaryons, the difference increases for decuplet
and is large for antidecuplet. By this reason the method of
calculation should be found where results do not depend on the
starting configuration. It can be, probably, some kind of 
"slow rotator" approximation used previously in $^{13,8b}$.

The inclusion of the configuration mixing into consideration $^{14}$
usually increases the mass splittings within multiplets, although it
does not change the results crucially.\\

5. The main uncertainty in the masses of all predicted states comes from
the poor known Casimir energies (CE) of states - the loop corrections 
of the order of $N_c^0$ to the classical masses of solitons. 
The CE was estimated for
the $B=1$ hedgehog $^{15}$ and also for $B=2$ $SO(3)$ hedgehog
$^{7}$. For $B=1$ case it has right sign and order of magnitude,
about $(-1  - -1.5)$ $Gev$. For the torus-like
 states $^{3}$ the $CE$ was not estimated yet.

The skyrmion molecules found in $^{10}$ should have the lowest 
uncertainty in Casimir energies relative to the $B=1$ states since 
in the molecule
unit skyrmions are only slightly deformed in comparison with starting
unperturbed configurations. Therefore, one can hope that the property of 
binding of dibaryons belonging to the lowest multiplets, octet and
decuplet, will not dissappear after inclusion of Casimir energy.

The prediction of the existence of multiplets of strange dibaryons
(also tribaryons, etc.), some of them being bound relative to strong 
interaction, remains a challenging property of the chiral soliton 
approach. This prediction is on the same level as the existence of
strange hyperons in the $B=1$ sector of the model because, within the
chiral soliton approach, skyrmions with different values of $B$ are
considered on equal footing. However, further
theoretical studies and comparison with predictions of other models
(see, e.g. $^{16}$) is of interest. 

Probably, we are standing now 
before the gate into "strange" world, the world of strange nuclear 
fragments, strangelets, etc. But this gate can be much more narrow 
than it was believed up till now. The realization of the Japan Hadron
Project could help much to find and to open this gate. \\
 
I am thankful to Bernd Schwesinger and Boris Stern for fruitful
collaboration. \\

\vglue 0.4cm
{\tenbf\noindent References}
\vglue 0.2cm

\end{document}